\documentclass[prl,superscriptaddress,twocolumn,showpacs,amsfonts,amsmath,a4paper]{revtex4}
\usepackage{graphicx}
\usepackage{dcolumn}
\usepackage{mathrsfs}
\usepackage{bm}

\usepackage{graphicx}
\usepackage{dcolumn}
\usepackage{bm}
\usepackage[usenames]{color}

\begin{document}

\title{Constraining the High-Density Behavior of Nuclear Symmetry Energy with the Tidal Polarizability of Neutron Stars}
\author{F. J. Fattoyev}
\email{Farrooh.Fattoyev@tamuc.edu} \affiliation{Department of
Physics and Astronomy, Texas A\&M University-Commerce, Commerce,
Texas 75429-3011, USA} \affiliation{Institute of Nuclear Physics,
Tashkent 100214, Uzbekistan}
\author{J. Carvajal}
\email{Jose.Carvajal002@mymdc.net} \affiliation{Department of
Physics and Astronomy, Texas A\&M University-Commerce, Commerce,
Texas 75429-3011, USA} \affiliation{Miami Dade College, 10525 SW
42nd TERR, Miami, Florida 33165, USA}
\author{W. G. Newton}
\email{William.Newton@tamuc.edu} \affiliation{Department of Physics
and Astronomy, Texas A\&M University-Commerce, Commerce, Texas
75429-3011, USA}
\author{Bao-An Li}
\email{Bao-An.Li@tamuc.edu} \affiliation{Department of Physics and
Astronomy, Texas A\&M University-Commerce, Commerce, Texas
75429-3011, USA} \affiliation{Department of Applied Physics, Xian
Jiao Tong University, Xian 710049, China}
\date{\today}
\begin{abstract}

Using a set of model equations of state satisfying the latest
constraints from both terrestrial nuclear experiments and
astrophysical observations as well as state-of-the-art nuclear
many-body calculations of the pure neutron matter equation of state,
the tidal polarizability of canonical neutron stars in coalescing
binaries is found to be a very sensitive probe of the high-density
behavior of nuclear symmetry energy which is among the most
uncertain properties of dense neutron-rich nucleonic matter.
Moreover, it changes less than $\pm 10\%$ by varying various
properties of symmetric nuclear matter and symmetry energy around
the saturation density within their respective ranges of remaining
uncertainty.

\end{abstract}

\maketitle

{\textit {Introduction.}}---Understanding the nature of neutron-rich
nucleonic matter is a major thrust of current research in both
nuclear physics and astrophysics~\cite{NSACLRP2007}. To realize this
goal, many experiments and observations are being carried out or
proposed using a wide variety of advanced new facilities, such as,
Facilities for Rare Isotope Beams (FRIB), X-ray satellites and
gravitational wave detectors. Most critical to interpreting results
of these experiments and observations is the equation of state (EOS)
of neutron-rich nucleonic matter, i.e., $E(\rho,\alpha)=E_0(\rho) +
S(\rho) \alpha^2 + \mathcal{O}(\alpha^4)$, where $E(\rho,\alpha)$
and $E_0(\rho)$ are the specific energy in asymmetric nuclear matter
of isospin asymmetry $\alpha=(\rho_n-\rho_p)/\rho$ and symmetric
nuclear matter (SNM), respectively, and $S(\rho)$ is the symmetry
energy encoding the energy cost of converting all protons in SNM to
neutrons. Thanks to the continuing efforts of both the nuclear
physics and astrophysics community over several decades, the EOS of
SNM around the saturation density $\rho_0$ has been well
constrained. Moreover, combining information from studying the
collective flow and kaon production in relativistic heavy-ion
collisions in several terrestrial nuclear physics
laboratories~\cite{Danielewicz:2002pu} and the very recent discovery
of the maximum mass of neutron stars~\cite{Demorest:2010bx}, the EOS
of SNM has been limited in a relatively small range up to about
$4.5\rho_0$. The symmetry energy $S(\rho)$ is a vital ingredient in
describing the structure of rare isotopes and their reaction
mechanisms. It also determines uniquely the proton fraction and thus
the cooling mechanism, appearance of hyperons and possible kaon
condensation in neutron stars. Moreover, it affects significantly
the structure, such as the radii, moment of inertia and the
core-crust transition density, as well as the frequencies and
damping times of various oscillation modes of neutron stars, see,
e.g., Refs.~\cite{Lattimer:2000kb,Lattimer:2004pg} for reviews.
Intensive efforts devoted to constraining $S(\rho)$ using various
approaches have recently led to a close merger around
$S(\rho_0)\approx 30$ MeV and its density slope $L\equiv 3 \rho_0
\left(d S(\rho)/d \rho\right)_{\rho_0}\approx 50$ MeV with a few
exceptions, although the error bars for $L$ from different
approaches may vary broadly~\cite{Xu:2010fh, Newton:2011dw,
Steiner:2011ft, Lattimer:2012xj, Dutra:2012mb, Tsang:2012se,
Fattoyev:2012ch}. On the other hand, the high-density behavior of
$S(\rho)$ remains very uncertain despite its importance to
understanding what happens in the core of neutron
stars~\cite{Kutschera:1993pe, Kubis:1999sc,
Kubis:2002dr,Wen:2009av,Lee:2010sw} and in reactions with high
energy radioactive beams~\cite{Li:2008gp}. The predictions for the
high-density behavior of the symmetry energy from all varieties of
nuclear models diverge dramatically~\cite{Brown:2000pd,
Szmaglinski:2006fz, Dieperink:2003vs}, with some models predicting
very stiff symmetry energies that increase continuously with
density~\cite{Steiner:2004fi, Lee:1998zzd, Horowitz:2000xj,
Dieperink:2003vs, Chen:2007ih, Li:2006gr}, and others predicting
relatively soft ones, or an $S(\rho)$ that first increases with
density, then saturates and starts decreasing with increasing
density~\cite{Pandharipande:1972, Friedman:1981qw, Wiringa:1988tp,
Kaiser:2001jx, Krastev:2006ii, Brown:2000pd, Chabanat:1997qh,
RikovskaStone:2003bi, Chen:2005ti, Decharge:1979fa, Das:2002fr,
Khoa:1996, Basu:2007ye, Myers:1994ek, Banik:1999cb,
Chowdhury:2009jn}. These uncertainties can be traced to our poor
knowledge about the isospin dependence of strong interaction in
dense neutron-rich medium, particularly the spin-isospin dependence
of three-body and many-body forces, the short-range behavior of
nuclear tensor force and the isospin dependence of nucleon-nucleon
correlations in dense medium, see, e.g. Refs.~\cite{Xu:2009bb,
Xu:2012hf}. Little experimental progress has been made in
constraining the high density $S(\rho)$ partially because of the
lack of sensitive probes. While several observables have been
proposed~\cite{Li:2008gp} and some indications of the high-density
$S(\rho)$ have been reported recently~\cite{Xiao:2009zza,
Russotto:2011hq}, conclusions based on terrestrial nuclear
experiments remain controversial~\cite{Trautmann:2012nk}. To our
best knowledge, the only astrophysical probe for high density
$S(\rho)$ proposed so far is the late time neutrino signal from a
core collapse supernova~\cite{Roberts:2011yw}. In this Letter, we
show that the tidal polarizability of canonical neutron stars in the
coalescing binaries is a very sensitive probe of the high-density
behavior of nuclear symmetry energy independent of the remaining
uncertainties of the SNM EOS and $S(\rho)$ near saturation density.

{\textit {Stellar structure and tidal polarizability.}}
---Coalescing binary neutron stars are among the most promising
sources of gravitational waves (GW). One of the most important
features of binary mergers is the tidal deformation neutron stars
undergo as they approach each other prior to merger, the strength of
which can give us precious information about the neutron-star matter
EOS \cite{Flanagan:2007ix, Hinderer:2007mb, Binnington:2009bb,
Damour:2009vw, Damour:2009wj, Hinderer:2009ca, Postnikov:2010yn,
Baiotti:2010xh, Baiotti:2011am, Lackey:2011vz, Pannarale:2011pk,
Damour:2012yf}. At the early stage of an inspiral tidal effects may
be effectively described through the tidal polarizability parameter
$\lambda$ \cite{Flanagan:2007ix, Hinderer:2009ca, Damour:2009vw,
Damour:2009wj} defined via $Q_{ij}= - \lambda \mathcal{E}_{ij}$,
where $Q_{ij}$ is the induced quadrupole moment of a star in binary,
and $\mathcal{E}_{ij}$ is the static external tidal field of the
companion star. The tidal polarizability can be expressed in terms
of the dimensionless tidal Love number $k_2$ and the neutron star
radius $R$ as $\lambda = 2k_2R^5/(3G)$. The tidal Love number $k_2$
is found using the following expression~\cite{Hinderer:2007mb,
Postnikov:2010yn}:
\begin{eqnarray} \nonumber
k_2 &=& \frac{1}{20}\left(\frac{R_s}{R}\right)^5 \left(1-
\frac{R_s}{R}\right)^2\left[2- y_{R} + \left(y_R-1\right)
\frac{R_s}{R} \right] \times \\
\nonumber &\times& \bigg\{\frac{R_s}{R} \bigg(6 - 3 y_R +
\frac{3R_s}{2R} \left(5y_R-8\right) +
\frac{1}{4}\left(\frac{R_s}{R}\right)^2 \bigg[26 -\\
\nonumber &-& 22y_R + \left(\frac{R_s}{R}\right) \left(3y_R-2\right)
+ \left(\frac{R_s}{R}\right)^2 \left(1+ y_R\right) \bigg]\bigg) +\\
\nonumber &+& 3\left(1-\frac{R_s}{R}\right)^2\left[2- y_{R} +
\left(y_R-1\right) \frac{R_s}{R} \right] \times \\
 &\times& \log \left(1- \frac{R_s}{R}\right)\bigg\}^{-1} \label{TidalLove} \ ,
\end{eqnarray}
where $R_s \equiv 2 M$ is the Schwarzschild radius of the star, and
$y_R \equiv y(R)$ can be calculated by solving the following
first-order differential equation:
\begin{eqnarray}
r \frac{d y(r)}{dr} + {y(r)}^2 + y(r) F(r) + r^2 Q(r) = 0
\label{TidalLove2} \ ,
\end{eqnarray}
with
\begin{eqnarray}
F(r) = \frac{r-4 \pi r^3 \left( \mathcal{E}(r) - P(r)\right) }{r-2
M(r)} \ ,
\end{eqnarray}
\begin{eqnarray}
\nonumber Q(r) &=& \frac{4 \pi r \left(5 \mathcal{E}(r) +9 P(r) +
\frac{\mathcal{E}(r) + P(r)}{\partial P(r)/\partial
\mathcal{E}(r)} - \frac{6}{4 \pi r^2}\right)}{r-2M(r)} - \\
&-&  4\left[\frac{M(r) + 4 \pi r^3
P(r)}{r^2\left(1-2M(r)/r\right)}\right]^2 \ .
\end{eqnarray}
The Eq. (\ref{TidalLove2}) must be integrated together with the
Tolman-Oppenheimer-Volkoff (TOV) equation. Given the boundary
conditions in terms of $y(0) = 2$, $P(0)\!=\!P_{c}$ and
$M(0)\!=\!0$, the tidal Love number can be obtained once an EOS is
supplied. Previous studies have used both polytropic EOSs and
several popular nuclear EOSs available in the literature~
\cite{Flanagan:2007ix, Hinderer:2007mb, Binnington:2009bb,
Damour:2009vw, Damour:2009wj, Hinderer:2009ca, Postnikov:2010yn,
Baiotti:2010xh, Baiotti:2011am, Lackey:2011vz,
Pannarale:2011pk,Damour:2012yf}. While other particles may be
present, for the purpose of this work, it is sufficient to assume
that neutron stars consist of only neutrons (n), protons (p),
electrons (e) and muons $(\mu)$ in $\beta$-equilibrium.

{\textit{ Constrained EOS of neutron-rich nuclear matter.}}
\begin{figure}
\includegraphics[width=1.0\columnwidth,angle=0]{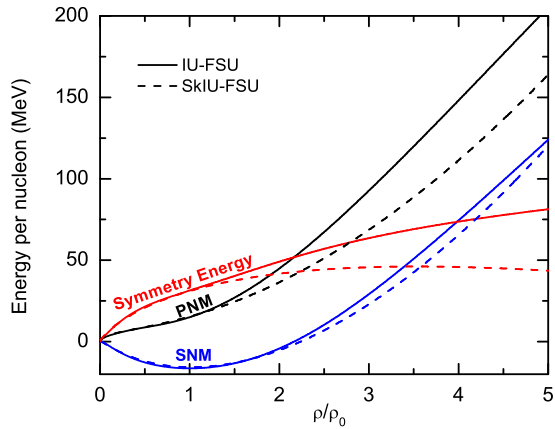}
\caption{(Color online) The EOS of SNM and PNM as well as the
symmetry energy as a function of density obtained within the IU-FSU
RMF model and the SHF approach using the SkIU-FSU parameter set.}
\end{figure}
---We use two classes of nuclear EOSs within the
Relativistic Mean Field (RMF) model and the Skyrme Hartre-Fock (SHF)
approach. All of these EOSs are adjusted to satisfy the following
four conditions within their respective uncertain ranges: (1)
reproducing the PNM EOS at sub-saturation densities predicted by the
latest state-of-the-art microscopic nuclear many-body
theories~\cite{Schwenk:2005ka, Friedman:1981qw, vanDalen:2009vt,
Hebeler:2009iv, Gandolfi:2008id, Gezerlis:2009iw, Vidana:2001ei};
(2) predicting correctly saturation properties of symmetric nuclear
matter, i.e, nucleon binding energy $B = -16 \pm 1 \, {\rm MeV}$ and
incompressibility $K_{0} = 230 \pm 20 \, {\rm MeV}$ and nucleon
effective mass $M^{\ast}_0 = 0.61\pm 0.03 \, M$ at saturation
density $\rho_0 = 0.155 \pm 0.01 \, {\rm fm}^{-3}$; (3) predicting a
fiducial value of symmetry energy  $S(2\rho_0/3) = 26 \pm 0.5 \,
{\rm MeV}$, $J \equiv S(\rho_0)=31 \pm 2 \, {\rm MeV}$ and the
density slope of symmetry energy $L = 50 \pm 10 \, {\rm MeV}$; (4)
passing through the terrestrial constraints on the EOS of SNM
between $2\rho_0$ and 4.5$\rho_0$~\cite{Danielewicz:2002pu} and
giving a maximum mass of neutron stars of about $2M_{\odot}$
assuming they are made of only the $npe\mu$ matter without
considering other degrees of freedom or invoking any exotic
mechanism~\cite{Demorest:2010bx,Steiner:2010fz}. As an example, two
such EOSs obtained using the IU-FSU RMF model~\cite{Fattoyev:2010mx}
and the SHF using the SkIU-FSU parameter set~\cite{Fattoyev:2012ch}
are shown in Fig.~1. By design, they both have the same EOS for SNM
and PNM around and below $\rho_0$. Thus, at sub-saturation densities
the values of $S(\rho)$ which is approximately the difference
between the EOSs for PNM and SNM are almost identical for the two
models. However, the values of $S(\rho)$ are significantly different
above about $1.5\rho_0$ with the IU-FSU leading to a much stiffer
$S(\rho)$ at high densities. More quantitatively, the $S(\rho)$ with
IU-FSU is $40-60\%$ higher in the density range of $\rho/\rho_0=3-4$
expected to reach in the core of canonical neutron stars.

\begin{figure}
\includegraphics[width=1.0\columnwidth,angle=0]{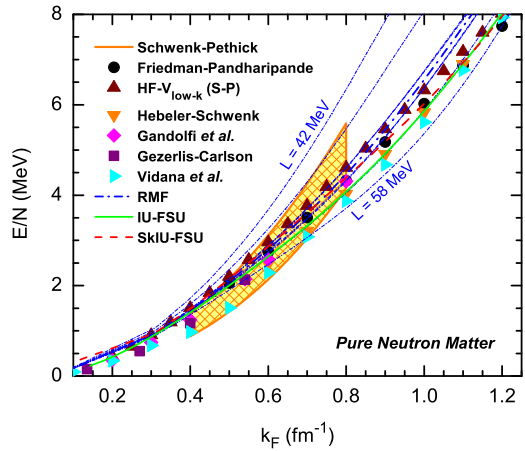}
\caption{(Color online) Energy per nucleon as a function of the
Fermi momentum for PNM for selected models described in the text.}
\end{figure}
\begin{figure}
\includegraphics[width=1.0\columnwidth,angle=0]{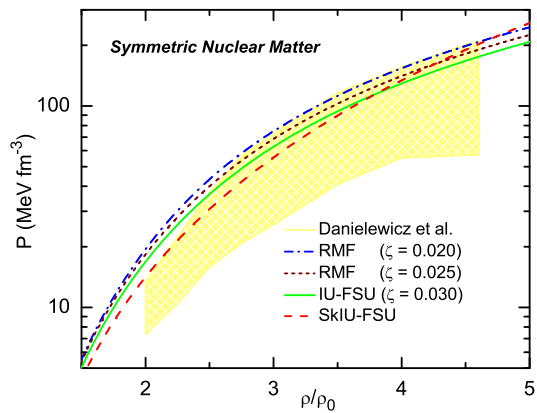}
\caption{(Color online) The pressure of SNM given as the function of
baryon density.  Here $\rho_0$ is the nuclear matter saturation
density and the shaded area represents the EOS extracted from the
analysis of \cite{Danielewicz:2002pu}.}
\end{figure}

To test the sensitivity of the tidal polarizability to variations of
properties of neutron-rich nuclear matter around $\rho_0$ within the
constraints listed above, we build 17 RMF parameterizations by
systematically varying the values of $K_0$, $M_{\rm 0}^{\ast}$, $L$,
and the $\zeta$ parameter of the RMF model that controls the
omega-meson self interactions~\cite{Mueller:1996pm} and subsequently
the high-density component of the EOS of SNM. Besides the
constraints listed above, all parameter sets can correctly reproduce
the experimental values for the binding energy and charge radius of
$^{208}$Pb and the ground state properties of other closed shell
nuclei within 2\% uncertainty~\cite{Todd:2003xs}. As a reference for
comparisons, we select $K_0 = 230$ MeV, $M_0^{\ast} = 0.61$ $M$, $L
=50$ MeV, and $\zeta = 0.025$ for our base model, which predicts
$\rho_0 = 0.1524$ fm$^{-3}$, $B = -16.33$ MeV and $J = 31.64$ MeV.
The representative model EOSs for PNM at sub-saturation densities
and those for SNM at supra-saturation densities are compared with
their constraints in Fig.~2 and Fig.~3, respectively. It is seen
that the SkIU-FSU and all the RMF models with $42< L< 58$ MeV can
satisfy the PNM EOS constraint. Also, they can all satisfy
simultaneously the high density SNM EOS constraint with $0.02<
\zeta< 0.03$. Moreover, they all give a maximum mass for neutron
stars between $1.94 M_{\odot}$ and $2.07 M_{\odot}$ and radii
between 12.33 km and 13.22 km for canonical neutron stars
\cite{Fattoyev:2010mx} consistent with existing
observations~\cite{Demorest:2010bx,Steiner:2010fz,Ozel:2010fw}.

{\textit {Results and discussions.}}
\begin{figure}
\includegraphics[width=1.0\columnwidth,angle=0]{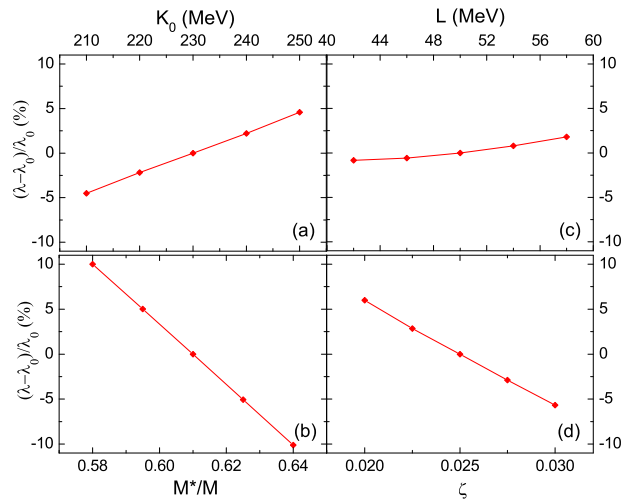}
\caption{(Color online) Percentage changes in the tidal
polarizability of a 1.4 solar mass neutron star by individually
varying properties of nuclear matter $K_0$ (a), $M^{\ast}$ (b), $L$
(c), and the $\zeta$ parameter (d) of the RMF model with respect to
the value using the base model.}
\end{figure}
\begin{figure}
\includegraphics[width=1.0\columnwidth,angle=0]{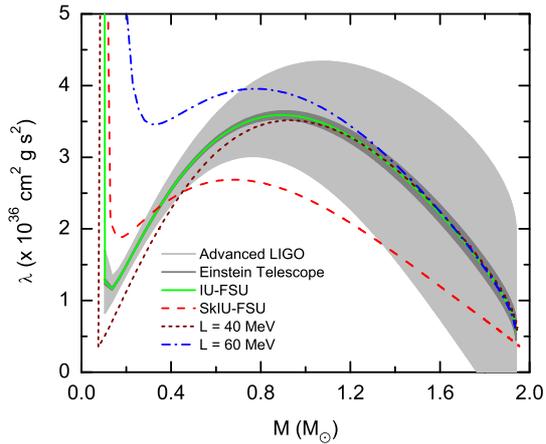}
\caption{(Color online) Tidal polarizability $\lambda$ of a single neutron star as
a function of neutron-star mass for a range of EOS that allow
various stiffness of symmetry energies. A crude estimate of
uncertainties in measuring $\lambda$ for equal mass binaries at a
distance of $D = 100$ Mpc is shown for the Advanced LIGO (shaded
light-grey area) and the Einstein Telescope (shaded dark-grey
area).}
\end{figure}
---First, we examine sensitivities of the tidal polarizability
$\lambda$ of a $1.4M_{\odot}$ neutron star to the variations of SNM
properties and the slope of the symmetry energy around $\rho_0$ in
Fig.~4. The changes of $\lambda$ relative to the values for our base
RMF model are shown for the remaining RMF EOSs. It is very
interesting to see that the tidal polarizability is rather
insensitive to the variation of $L$ although it changes up to $\pm
10\%$ with $K_0$, $M^{\ast}$ and $\zeta$ within their individual
uncertain ranges. While the averaged mass is $M = 1.33 \pm 0.05 \,
M_{\odot}$, neutron stars in binaries have a broad mass
distribution~\cite{Ozel:2012ax}. It is thus necessary to investigate
the mass dependence of the tidal polarizability. Whereas what can be
measured for a neutron star binary of mass $M_1$ and $ M_2$ is the
mass-weighted polarizability $\tilde{\lambda} =
\left[\lambda_1\left(M_1+12 M_2\right)/{M_1}  + 1 \leftrightarrow 2
\right]/26$~\cite{Hinderer:2009ca}, for the purpose of this study it
is sufficient to consider binaries consisting of two neutron stars
with equal masses. What can we learn from the tidal polarizability
of light and massive neutron stars, respectively? Shown in Fig.~5
are the tidal polarizability $\lambda$ as a function of neutron-star
mass for a range of EOSs. Most interestingly, it is seen that the
IU-FSU and SkIU-FSU models which are different only in their
predictions for the nuclear symmetry energy above about $1.5\rho_0$
as shown in Fig.~1 lead to significantly different $\lambda$ values
in a broad mass range from 0.5 to 2 $M_{\odot}$. More
quantitatively, a $41\%$ change in $\lambda$ from $2.828\times
10^{36}$ (IU-FSU) to $1.657\times 10^{36}$ (SkIU-FSU) is observed
for a canonical neutron star of 1.4 $M_{\odot}$. For a comparison,
we notice that this effect is as strong as the symmetry energy
effect on the late time neutrino flux from the cooling of
proto-neutron stars~\cite{Roberts:2011yw}. Moreover, it is shown
that the variation of $L$ has a very small effect on the tidal
polarizability $\lambda$ of massive neutron stars, which is
consistent with the results shown in Fig.\ 4. On the other hand, the
$L$ parameter affects significantly the tidal polarizability of
neutron stars with $M\leq 1.2M_{\odot}$. These observations can be
easily understood. From Eq. (\ref{TidalLove}) the Love number $k_2$
is essentially determined by the compactness parameter $M/R$ and the
function $y(R)$. Both of them are obtained by integrating the EOS
all the way from the core to the surface. Since the saturation
density approximately corresponds to the central density of a
$0.3M_{\odot}$ neutron star, one thus should expect that only the
Love number of low-mass neutron stars to be sensitive to the EOS
around the saturation density. However, for canonical and more
massive neutron stars, the central density is higher than
$3-4\rho_0$, and therefore both the compactness $M/R$ and $y(R)$
show stronger sensitivity to the variation of EOS at
supra-saturation densities. Since all the EOSs for SNM at
supra-saturation densities have already been constrained by the
terrestrial nuclear physics data and required to give a maximum mass
about $2M_{\odot}$ for neutron stars, the strongest effect on
calculations of the tidal polarizability of massive neutron stars
should therefore come from the high-density behavior of the symmetry
energy.

It has been suggested that the Advanced LIGO-Virgo detector may
potentially measure the tidal polarizability of binary neutron stars
with a moderate accuracy. Are the existing or planned GW detectors
sensitive enough to measure the predicted effects of high-density
symmetry energy on the tidal polarizability? To answer this
question, as an example we estimate uncertainties in measuring
$\lambda$ for equal mass binaries at an optimally-oriented distance
of $D = 100$ Mpc~\cite{Hinderer:2009ca, Abadie:2010cf} using the
same approach as detailed in Refs.~\cite{Hinderer:2009ca,
Damour:2012yf}. These are shown for the Advanced LIGO-Virgo (shaded
light-grey area) and the Einstein Telescope (shaded dark-grey area)
in Fig. 5.  It is seen that the Advanced LIGO-Virgo will unlikely
constrain the EOS and symmetry energy at supra-saturation densities
within the estimated uncertainty, although it is possible that a
rare but nearby binary system may be found and provide a much more
tighter constraint~\cite{Hinderer:2009ca}. Nevertheless,
measurements for binaries consisting of light neutron stars can
still help further constrain the symmetry energy around the
saturation density. On the other hand, it is exciting to see that
the narrow uncertain range for the proposed Einstein Telescope will
enable it to tightly constrain the symmetry energy especially at
high densities.

{\textit {Conclusions.}}---Using the EOSs for neutron-rich nucleonic
matter satisfying the latest constraints from both terrestrial
nuclear experiments and astrophysical observations, as well as the
state-of-the-art nuclear many-body calculations for PNM EOS, we
found that the tidal polarizability of canonical neutron starts in
coalescing binaries is very sensitive to the high-density behavior
of nuclear symmetry energy, but little affected by the variations of
SNM EOS and symmetry energy around the saturation density within
their remaining uncertainty ranges. Future measurements of the tidal
polarizability of neutron stars using the proposed Einstein
Telescope will help constrain stringently the high-density behavior
of nuclear symmetry energy, and thus the nature of dense
neutron-rich nucleonic matter.

{\textit {Acknowledgments.}} ---We would like to thank Dr. J.
Piekarewicz and Dr. Jun Xu for various useful discussions. This work
is supported in part by the National Aeronautics and Space
Administration under grant NNX11AC41G issued through the Science
Mission Directorate, and the National Science Foundation under
Grants No. PHY-0757839, No. PHY-1062613 and No. PHY-1068022.

\bibliography{ReferencesFJF}


\end{document}